\documentclass[aps,prl,showpacs,amsmath,amssymb,
	preprint,onecolumn]{revtex4}

\usepackage[dvips]{graphicx}
\usepackage{dcolumn}
\usepackage{bm}
\usepackage{epsf}

\begin{document}

\title{Complete chaotic
synchronization in mutually coupled time-delay systems}

\author{Alexandra S. Landsman$^1$  and Ira B. Schwartz$^1$}
\affiliation{$^1$US Naval Research Laboratory, Code 6792, Nonlinear Systems
  Dynamics Section, Plasma Physics Division, Washington, DC 20375}
\email{<alandsma@cantor.nrl.navy.mil>}

\begin{abstract}
Complete chaotic synchronization of end lasers
has been observed in a line of mutually coupled, time-delayed
system of three lasers, with no direct communication between the
end lasers.
The present paper uses ideas from generalized synchronization to explain the
complete synchronization in the presence of long
coupling delays, applied
to a model of  mutually coupled semiconductor lasers in a line.
These ideas significantly simplify the analysis by casting the
stability in terms of the local dynamics of each laser.
The variational
equations near the synchronization manifold are analyzed, and used to derive
the synchronization condition that is a function of parameters.  The results
explain and predict
the dependence of synchronization on various parameters, such as
time-delays, strength of coupling and dissipation.  The ideas can be applied
to understand complete synchronization in other chaotic systems with
coupling delays and no direct communication between synchronized
sub-systems.    
\end{abstract}

\pacs{05.45.Xt, 05.45.-a, 42.65.Sf, 05.45.Vx}

\maketitle

\section{\label{sec:intro} I. Introduction}

Synchronized chaotic oscillations have been found in many nonlinear
systems, from lasers \cite{MirassoVCMP04} to neural networks \cite{CiszakCMMT03}. Many types of synchronization have been observed, including complete
synchronization, phase-locking, and generalized synchronization, in the case
of uni-directionally coupled systems. For an overview of the large body of
work done on synchronization, and its sub-classes see for example 
\cite{BoccalettiKOVZ02,pikovsky}.  While extensive work has been done on mutually
coupled systems, general analytic techniques 
for analyzing chaotic synchronization in 
time-delay mutually  coupled systems have not been well
developed. Essentially, there exist two methods for analyzing
synchronization in coupled systems: A Lyapunov function approach
\cite{Hale:1971:FDE} and a master stability approach \cite{PecoraC98}.
Time delays considerably 
complicate the analysis, possibly introducing infinite degrees of freedom,
and  resulting in new types of dynamics \cite{Hale:1971:FDE}.
The present paper proposes an approach for understanding and predicting
chaotic synchronization of time-delayed mutually coupled systems, possessing
internal symmetry.


In the case of internal symmetry in a system, where the equations of motion are
invariant with respect to interchange of some variables, there is a solution
where these variables are exactly equal \cite{TerryTDVZAR99}.  For example, for the
case of three nonlinear
oscillators coupled in a line, 
if the equations of the outside oscillators
are identical, then given the same initial conditions, these oscillators
will have identical dynamics, including the possibility of  chaotic solutions.
In that case, the system can be reduced
to two coupled oscillators.  
If the symmetric solution is asymptotically stable to perturbations off the synchronization manifold,  then the dynamics of outside oscillators are
synchronized.  Thus the requirement for synchronization in the long
time limit is that the largest
Lyapunov exponent, with respect to perturbations transverse to the
synchronization manifold, is negative, resulting in decay of the initial
perturbation back to synchronized state \cite{PecoraC98}.
In general, Lyapunov exponents have to be calculated numerically. However, as
will be shown, an analytic estimate can be made in some cases by linearizing
about the synchronous state.  

\begin{figure}[ht]
\hspace*{-1 cm}
{
\epsfxsize=6in
\epsffile{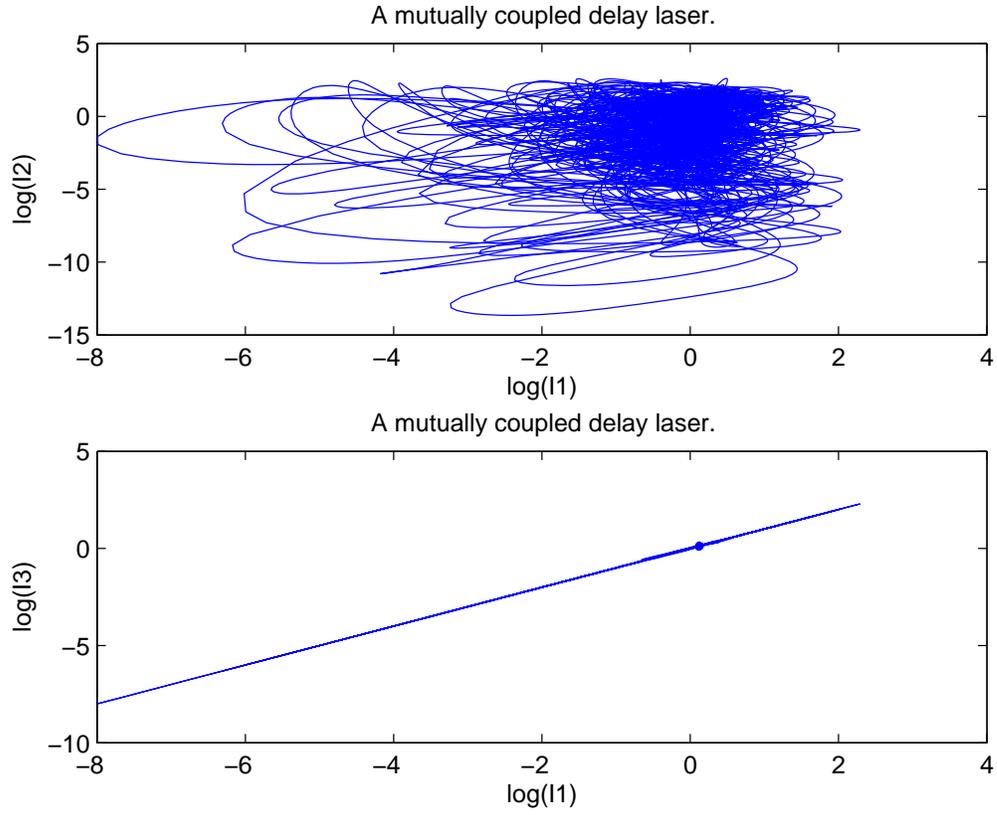}
}
\caption{Top: Intensity of Laser 1 vs. Laser 2.  Bottom:
Laser 1 vs. Laser 3.  Straight line indicates 
complete synchronization of outer lasers.
$\tau=30$, $\epsilon=\sqrt{0.001}$, $\delta_1 = \delta_2 = 6.5 \epsilon$ }
\label{fig:lasers_synch}
\end{figure}

Although there  is extensive work on synchronization of coupled
systems, studies of chaotic 
synchronization in time-delayed systems is
much less extensive by comparison.  Some analysis exists on synchronization of
coupled semiconductor lasers without delays \cite{TerryTDVZAR99,CarrTS06}.  
However, it remains to be
explained, for example, 
why in a coupled three laser system, outer lasers show complete
synchronization, in the presence of long time delays (compared to the
internal dynamics of each laser) and no
communication except via the middle laser, which itself is not
synchronized with the end lasers \cite{FishcervbpMtG06}. (See Figure
\ref{fig:lasers_synch} below for an example).
The present paper aims to explain this phenomena observed in lasers
and other time-delayed systems using ideas from generalized
synchronization.  The paper is organized as follows:  In Section II, the
general equations for the 
three subsystems coupled by nearest neighbor interactions
with delays are introduced, and the equations are linearized close to the
synchronization manifold, using internal symmetry.  
Section III uses synchronization ideas developed in Section II to
explain complete synchronization of semiconductor lasers in the presence of
long delays and with no direct communication between the outer lasers.
Section IV concludes and summarizes.

\section{\label{sec:basic} II. Long delays and generalized synchronization}

\begin{figure}[ht]
\hspace*{-1 cm}
{
\epsfxsize=6in
\epsffile{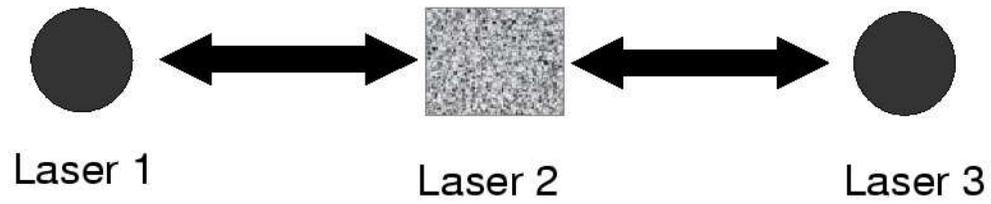}
}
\caption{A schematic showing how three lasers are coupled in a
  line. The outer two lasers (circles) are identical, while the middle
  laser (square) is
  detuned from the rest. }
\label{fig:3lasers}
\end{figure}

We deal with mutually coupled, oscillatory
time-delay systems possessing internal symmetry,
with respect to interchange of some variables.  The system can be broken up
into three coupled parts as depicted in Fig. \ref{fig:3lasers}: the ``center'' and two identical sub-parts, that
possess symmetry with respect to interchange of variables:

\begin{eqnarray}
\frac{d \mathbf{z_1}}{dt} & = & \mathbf{F}\left(\mathbf{z_1}(t)\right) 
+ \delta_1 \cdot \mathbf{G} \left(\mathbf{z_2} \left(t-\tau \right) \right) \\
\frac{d \mathbf{z_2} }{dt} & 
= & \mathbf{\tilde F} \left(\mathbf{z_2}(t)\right) 
+ \delta_2 \cdot \mathbf{\tilde G} \left(\mathbf{z_1} \left(t-\tau \right), \mathbf{z_3} \left(t-\tau \right) \right) \\
\frac{d \mathbf{z_3}}{dt} & = & \mathbf{F} \left(\mathbf{z_3}(t)\right) 
+ \delta_1 \cdot \mathbf{G}\left(\mathbf{z_2} \left(t-\tau \right) \right)
\label{eq:z}
\end{eqnarray}
where $\mathbf{z_i}$ are vector variables of some dimension $M$ and $N$,
for the outer and the middle subsystem, respectively.  
Variables $\mathbf{z_1}$ and $\mathbf{z_3}$ 
are symmetric with respect to interchange
of variables, while the center system,  $\mathbf{z_2}$  
may have different internal dynamics,
given by $\mathbf{\tilde F}$, and different coupling function 
$\mathbf{\tilde G}$.  
The delay in the coupling terms is fixed and  given by $\tau$, and
the strength of coupling from the center to the outer identical subsystems by
$\delta_1$, while from the outer to the center by $\delta_2$.  
Due to internal symmetry of the system, there exists an
identical solution for the outside subsystems: 
$\mathbf{z_1}(t) = \mathbf{z_3}(t) = \mathbf{\phi}(t)$.   
If the conditional Lyapunov exponents calculated with respect to perturbation
out of the synchronization manifold, $\mathbf{z_1}(t) = \mathbf{z_3}(t)$ 
are all negative, then the two outer subsystems are
synchronized.  This type of behavior where the two systems show identical
dynamics, even chaotic ones, is called complete synchronization.  
Figure \ref{fig:lasers_synch}
shows an example of complete synchronization in the case of a mutually
coupled three laser system using the model in Section III \cite{note3}.  
Calculating Lyapunov exponents is in
general complicated due to the presence of time-delays in the equations.  
The coupling term containing
delays, however, drops out if Eqs. (1), (3) are linearized about the
synchronous state: $\mathbf{z_1}(t) = \mathbf{z_3}(t)$.  
To study the stability of the symmetric solution, we introduce new variables:
$\mathbf{\triangle z_1}(t) =\mathbf{z_1}(t) - \mathbf{\phi}(t)$ and 
$\mathbf{\triangle z_3}(t) =\mathbf{z_3}(t) - \mathbf{\phi}(t)$.
Linearizing transverse to the synchronization manifold, we have:
\begin{equation}
\frac{d \mathbf{\triangle z_1}(t)}{dt} 
= \mathbf{J} \cdot \mathbf{\triangle z_1}(t)
\label{eq:lin}
\end{equation}
\begin{equation}
\frac{d \mathbf{\triangle z_3}(t)}{dt} 
= \mathbf{J} \cdot \mathbf{\triangle z_3}(t)
\label{eq:lin2}
\end{equation}
where $\mathbf{J}$ is the $\mathbf{M} \times \mathbf{M}$
Jacobian matrix of partial derivatives evaluated 
at $\mathbf{\phi}(t)$,
\begin{equation}
\mathbf{J}  = \frac{\partial \mathbf{F}\left(\mathbf{\phi}(t)
  \right)}{\partial \mathbf{z}}.
\label{eq:Jacobian}
\end{equation}
   
Here $\mathbf{\phi}(t)$ is the synchronous state that is determined by the dynamics
of Eqs. (1)-(3) and the initial conditions defined on $[-\tau,0)$.  Although the time delays and
dependence on $\mathbf{z_2}$ 
drop out of Eqs. (\ref{eq:lin}) and (\ref{eq:lin2}), 
they are involved implicitly in
determining $\mathbf{\phi}(t)$, the synchronization manifold.  
Equations (\ref{eq:lin}) and (\ref{eq:lin2}) are $M$-dimensional and therefore
have $M$ transverse Lyapunov exponents.  The largest of them determines the
stability of the transverse perturbation.
So that the
synchronized state $\mathbf{z_1}(t) = \mathbf{z_3}(t) = \mathbf{\phi}(t)$ 
is asymptotically
stable, if $\mathbf{\triangle z_{1,3}}(t) \rightarrow 0$ 
as $t \rightarrow \infty$ or if
all of the Lyapunov exponents in the linearized equations are negative.  

For $\delta_2 =0$, in Eq. (2), the dynamics of $\mathbf{z_{1,3}}$ become
that of a driven system, with $\mathbf{z_2}$ 
acting as the driver.  Then, the synchronized
dynamics correspond to generalized synchronization \cite{RulkovSTA95} whereby 
the driven subsystem becomes a function of the driver: 
$\mathbf{\phi}(t) = \mathbf{f(\Phi)}$.  
Here, $\mathbf{\Phi}$ are the dynamics of the driver obtained by integrating
Eq. (2), with $\delta_2 =0$:  $d \mathbf{\Phi}(t)/dt =  \tilde F
\left(\mathbf{\Phi}(t)\right)$.

While the exact form of the function between the driver and the driven systems
can be rather complicated and 
difficult to obtain, its existence can be inferred from the
synchronization of identical systems when started from different initial
conditions and being exposed to the same drive.  This method of detecting
generalized synchronization using identical driven systems is known as the
auxiliary systems approach \cite{AbarbanelRS96}.  
In order for the driven subsystems, $\mathbf{z_{1,3}}$ to
become synchronized, their dependence on initial
conditions has to "wash out" as a function of time.  This is due to the fact
that dependence on initial conditions prevents synchronization
by making the dynamics of $\mathbf{z_1}$ different from the dynamics
of $\mathbf{z_3}$.  This "washing out" of initial conditions is provided
by the dissipation in the system, which must therefore be either
present in the coupling term, or in the uncoupled dynamics of the
system itself. This can be seen by taking the
sum of Lyapunov exponents, 
for Eqs. (\ref{eq:lin}) and (\ref{eq:lin2}), which   
is related to contraction or expansion of
the phase-space volume of the dynamics transverse to the synchronization
manifold \cite{Hartman:1982:ODE},
\begin{equation}
\sum_{j=1}^M \lambda_j = \lim_{t\to\infty} \frac{1}{t} 
\ln|\det\left(\mathbf{\Psi(\triangle \mathbf{z_{1,3}})}\left(t\right)\right)|,
\label{eq:LyapSum}
\end{equation}
where $\mathbf{\Psi}$ is the fundamental matrix solution to
Equations (\ref{eq:lin}) and (\ref{eq:lin2}).
From Eq. (\ref{eq:LyapSum}), negative 
Lyapunov exponents sum corresponds to a contraction of phase-space of 
$\mathbf{\triangle z_{1,3}}$ dynamics as a function of time.  
In the case of synchronization in driven identical
systems, all Lyapunov exponents transverse to the synchronous
solution must be negative.  Thus there is a 
contraction of phase-space to a single
trajectory that is a function of the dynamics of the driver, $\mathbf{\Phi}$.
This shrinking of phase-space is either caused by dissipative coupling or
dissipation in the driven systems themselves.
The effect of dissipation on synchronization can be illustrated using
the case of a simple driven system,
\begin{equation}
\frac{dx}{dt} = -\epsilon x + \Phi(t)  \qquad \qquad \frac{dy}{dt} = -\epsilon y + \Phi(t)
\label{eq:xy}
\end{equation}
taking the difference between the two variables, $\triangle = x-y$, we get,
$d \triangle/dt = -\epsilon \triangle$,
where $\Phi(t)$ in Eq. (\ref{eq:xy}) is the 
chaotic signal provided by the driver (it can also
be a noisy signal) \cite{pikovsky,GuanLLG06}.  
Thus the difference in 
initial conditions between $x$ and $y$ 
decreases at the rate of dissipation, $\epsilon$,
leading to synchronization for large times.

In mutually coupled systems,  $\delta_2 \neq 0$, 
the dynamics of $\mathbf{z_2}$ 
are affected by $\mathbf{z_1}$ and $\mathbf{z_3}$.  In this case, the
synchronized state, $\phi(t)$, may depend on the 
initial conditions of all of the
three subsystems, $\{\mathbf{z_1}, \mathbf{z_2}, \mathbf{z_3}\}$, 
so that $\phi(t)$ can not be the result
of generalized synchronization, in a strict sense.  However, it takes a
time interval of $2 \tau$ for any change in the dynamics of systems 
$\mathbf{z_{1,3}}$ 
to affect the motion of these systems via mutual coupling.  
During this time interval of length, $2 \tau$, $\mathbf{z_{1,3}}$ 
can be viewed as driven by
$\mathbf{z_2}$, since the signal $\mathbf{z_{1,3}}$  
receives during that time-interval is not affected
by its dynamics on that interval.  Therefore, we examine the dynamics
in a time period on the order of the delay time , $\tau$.

The initial dynamics of $\mathbf{z_{1,3}}(t)$
on the time interval 
$t_0 \leq t < t_0 + \tau$, affect the dynamics of $\mathbf{z_{1,3}}(t)$ on
the time interval $t_0+2 \tau \leq t < t_0+3\tau$ 
via the mutual coupling term,
$\mathbf{G}\left(\mathbf{z_2}\left(t-\tau \right)\right)$ (see Eqs. (1) and (3)).
For chaotic systems,  the trajectories are
not correlated in time, and we assume
\begin{equation}
\langle \mathbf{z_i}(t) \cdot \mathbf{z_i}(t-t_0) \rangle \approx 0
\label{eq:corr}
\end{equation}
for $i=\{1,2,3\}$ and $t_0$ significantly
longer than the average period of oscillation.  
The above equation is true, in general, for non-periodic oscillations.
This can be seen by expanding the signal in a Fourier series:
$z_i (t) = \sum_{n=-\infty}^\infty A_n cos \left(n t + \phi \right)$.
Then, if there is a significant periodic component in $z_i (t)$ of
amplitude $A_n$, Eq. (\ref{eq:corr}) will be proportional to $A_n^2$
whenever $t_0$ is a multiple of $2 \pi/n$.  Thus, for example, Eq.
(\ref{eq:corr}) may not hold if the amplitude of an optoelectronicly
coupled laser (to be discussed in the following section) is too close
to the threshold, where the behavior can be approximated as coupled
linear oscillators (see Eqs.  (\ref{eq:lasers1})-(\ref{eq:lasers})), 
leading to a significant regular oscillatory component in the signal.
In this case, there may be a resonant interaction between the lasers,
which is sensitive to the specific value of the coupling delay, $\tau$.
For chaotic dynamics, we can substitute  $t_0=2 \tau$ into
Eq.  (\ref{eq:corr}), where $\tau$ is the delay, to see that there is no
significant correlation between the dynamics on that time scale, so that over
the round-trip time of $2 \tau$, 
the identical sub-systems $\mathbf{z_{1,3}}$ can be viewed as driven 
by some uncorrelated chaotic signal coming from $\mathbf{z_2}$. 
This assumption of a driver 
is only strictly valid on the time interval within the
round-trip time, since for longer time intervals the initial conditions of the
outer lasers at the beginning of the interval will affect their dynamics, via
the middle system, at a later time within the interval.

By perturbing the dynamics of the  outer subsystems from the  
synchronized state, it can be shown that
complete synchronization of the end subsystems in the presence of
long delays is similar to generalized synchronization, where the middle 
subsystem acts as the driver for the outer ones.
After the symmetric subsystems 
synchronize, $\mathbf{z_{1,3}}(t)=\mathbf{\phi}(t)$, 
one of them can be suddenly perturbed from its
symmetric state to an arbitrary position in phase-space at some $t=t_1$.  
In that case, the
perturbed system, as well as the unpertubed one, will receive the exact same 
signal from $\mathbf{z_2}$ as before, for $t < t_1 + 2 \tau$.  If the systems
synchronize again at some point during $t < t_1 + 2 \tau$, we will again have
$\mathbf{z_{1,3}}(t)=\mathbf{\phi}(t)$, 
where $\mathbf{\phi}(t)$ has not been affected by 
the perturbation during that time interval.
Thus the synchronized state, $\mathbf{\phi}(t)=\mathbf{z_{1,3}}(t)$ is 
clearly independent of perturbations
of subsystems $\mathbf{z_{1,3}}$ and must therefore be some
function of the middle subsystem, $\mathbf{z_2}$.  
This however is the same as what
happens in generalized synchronization, with the difference that the
trajectory of $\mathbf{z_2}$ itself may be affected by the initial starting 
conditions of the symmetric subsystems.
The following section uses a system of semiconductor lasers as an example for
application of these ideas to understand complete synchronization of the end
lasers in a three laser system.  

\section{\label{sec:lasers} III. Synchronization of semiconductor lasers with delays}

The discussion of the previous section can be applied to the study of
synchronization of a three laser system with delays.  A schematic diagram is
shown in Fig. (\ref{fig:3lasers}), where the outer lasers are identical, while
the middle laser is  detuned.

  The scaled equations of coupled
semiconductor lasers have the following form \cite{KimRACS05,carrS06}:
\begin{displaymath}
\frac{dy_1}{dt} = x_1 \left(1+y_1\right)
\end{displaymath}  
\begin{equation}
\frac{dx_1}{dt} = -y_1 - \epsilon x_1\left(a_1 + b_1 y_1\right) + \delta_2
y_2\left(t-\tau\right)
\label{eq:lasers1}
\end{equation}  
\begin{displaymath}
\frac{dy_2}{dt} = \beta x_2 \left(1+y_2\right) 
\end{displaymath}  
\begin{equation}
\frac{dx_2}{dt} = \beta \left(-y_2 
- \epsilon \beta x_2 \left(a_2 + b_2 y_2\right) \right)
+ \delta_1 \left(y_1\left(t-\tau\right) + y_3\left(t-\tau\right)\right)
\label{eq:lasers2}
\end{equation}
\begin{displaymath} 
\frac{dy_3}{dt} = x_3 \left(1+y_3\right) 
\end{displaymath}
\begin{equation}
\frac{dx_3}{dt} = -y_3 - \epsilon x_3\left(a_1 + b y_1\right) + \delta_2
y_2\left(t-\tau\right)
\label{eq:lasers}
\end{equation} 
Eqs. (\ref{eq:lasers1})-(\ref{eq:lasers}) have the same form as 
Eqs. (1)-(3) with $\mathbf{z_i} =
\{y_i, x_i\}$, where $i=1,3$ and $i=2$ for the outer and middle lasers,
respectively. Variables $y_i$ and $x_i$ denote scaled intensity and inversion
of the $ith$ laser, $\{a_1,a_2,b_1,b_2\}$ are loss terms, and $\epsilon$ is
the dissipation. (See \cite{SchwartzE94} for details of the derivation
form the original physical model.)  The above equations are coupled via
laser intensities, $y_i$, using optoelectronic incoherent 
coupling that does not contain phase information, unlike the coherent coupling
in Fischer et al \cite{FishcervbpMtG06}.  
Previously, the dynamics of two electronically coupled lasers have 
been explored \cite{Vicente04}, showing lag synchronization for
the case of two lasers, and isochronal synchronization if feedback is
added \cite{Vicente06}.  The above equations are scaled so that the relaxation
frequency is equal to one.  In the typical experimental set-up, the relaxation
oscillations are on the order of $2-3$ ns.  Since the delay time is scaled
by the relaxation frequency, and is at least an order of magnitude higher (for
long delays), a
typical delay time used in simulations could be about $\tau = 60$, which 
corresponds to about $20-30$ ns.

In the absence of dissipation, the uncoupled
system, $\delta_1=\delta_2=0$, is a nonlinear conservative system, with behavior similar to a
simple harmonic oscillator for small amplitudes, and becoming more pulse-like
at high amplitudes \cite{SchwartzE94}.  Dissipation, however, leads to energy loss, so that in
the absence of coupling between lasers, the dynamics would settle into a
steady state.  Thus mutual coupling acts like a drive by pumping energy into
the system.  For most cases, it can be assumed that dissipation is small:
$\epsilon \ll 1$.  Detuning of the middle laser from the outer ones is given by
$\beta$. 

The system described by Eqs. (\ref{eq:lasers1})-(\ref{eq:lasers}) shows complete synchronization
of outer lasers over a whole range of parameters.  
Figure \ref{fig:lasers_synch} shows that while the outer lasers can
become completely synchronized, there may be no apparent correlation between
the middle and the outer lasers.  
Since the outer lasers are identical, there is a solution of
Eqs. (\ref{eq:lasers1}) - (\ref{eq:lasers}) 
where $y_1 = y_3=Y(t)$ and $x_1=x_3=X(t)$.  
In this case,  Eqs.  (\ref{eq:lasers1}) -(\ref{eq:lasers}) 
reduce to four differential equations.  The
solution $y_1 = y_3$ and $x_1=x_3$ is stable if the Lyapunov exponents
transverse to the synchronization manifold are negative.  To investigate the
stability of the synchronized state we linearize about the synchronous
solution $\mathbf{\phi}(t) = \{X(t), Y(t)\}$.  Applying Eqs. (\ref{eq:lin})-
 (\ref{eq:Jacobian}) to Eqs. (\ref{eq:lasers1}) and  (\ref{eq:lasers}), we
get 
\begin{equation}
\begin{pmatrix} \dot{\triangle x_{1,3}} \\ \dot{\triangle y_{1,3}} \end{pmatrix} =
\begin{pmatrix} -\epsilon \lbrack a_1 + b_1 Y(t) \rbrack 
& -\lbrack 1+\epsilon b_1 X(t) \rbrack \\ 1+Y(t) & X(t) \end{pmatrix} \cdot 
\begin{pmatrix} \triangle x(t)_{1,3} \\ \triangle y(t)_{1,3} \end{pmatrix}
\label{eq:matrix}
\end{equation}
where $\{\triangle x(t)_{1,3}, \triangle y(t)_{1,3} \}$ are perturbations of
outer oscillators from the synchronous state $\{ X(t), Y(t) \}$.  This
synchronous state is obtained by starting the outer oscillators from the same
initial conditions and perturbing the system at some time, $t=t_1$.  The
perturbation will not affect the coefficient matrix in Eq. (\ref{eq:matrix})
until $t \geq t_1 + 2 \tau$.  So that in the time interval of $2 \tau$ the
dynamics off the synchronization manifold can be viewed as driven by an
uncorrelated chaotic signal $\{X(t), Y(t)\}$.  We can now apply Abel's
formula \cite{Hartman:1982:ODE}
to Eq. (\ref{eq:matrix}), which relates the Wronskian of the linearized system
to the trace of the matrix \cite{zill}.  Dropping the subscripts on linearized variables,
we get,
\begin{equation}  
W\left(t\right) = \det \begin{vmatrix} \triangle x & \triangle y
\\ \dot{\triangle x} & \dot{\triangle y} \end{vmatrix} 
= \exp\left( \int^t_{t_1} \{X(s) - \epsilon \left(a_1 + b_1 Y(s) \right) \} \cdot ds \right)
\label{eq:W}
\end{equation}
The Wronskian gives the phase-space volume dynamics  of the system $\{\triangle
x(t), \triangle y(t) \}$.  Equation (\ref{eq:W}) is valid over the
integration interval of twice the delay: $t_1 < t < t_1 + 2 \tau$.
This is due to the fact that it takes a time interval
of $2 \tau$ for a perturbation
in the outer laser to affect its dynamics via mutual coupling from the middle
laser.  Thus, during the time interval of $2 \tau$, the perturbed system acts
like a driven system in that its dynamics do not affect the signal it
receives, and therefore do not change the synchronized state, $\{X, Y\}$, 
making it independent of  $\{\triangle x, \triangle y \}$ 
dynamics over the integration interval.
  
Since the variable $Y(t)$ is the scaled intensity of the laser, from 
Eqs. (\ref{eq:lasers1})-(\ref{eq:lasers}), its minimum possible value is
$-1$.  Thus for $a_1 > b_1$ (a typical case), the contribution of the
dissipation term to the Wronskian is always negative.
The variable $X(t)$, on the other hand, is symmetric about zero, and
thus averages out to zero when integrated over a single period of oscillation.
It follows that if the integral in Eq. (\ref{eq:W}) is taken just over
a single oscillation of the laser, we get
\begin{equation}
\int^{t_1+T}_{t_1} \{X(s) - \epsilon \left(a_1 + b_1 Y(s) \right) \} \cdot ds
= - \epsilon \left(a_1 + b_1 \bar Y \right) T < 0
\label{eq:insert}
\end{equation}
where $T$ is the period of a single oscillation, and $\bar Y$ is the average
of $Y$ over a single period (unlike $X$, the $Y$ variable is not
symmetric about zero, which can readily be seen in the pulse-like
fluctuations of lasers at high intensities).  
It follows, that $X(s)$ in Eq. (\ref{eq:W}) 
averages out to zero if the integral is done over many periods of oscillation,
while the dissipation term, multiplied by $\epsilon$, provides a continuous
negative component.  If that continuous negative component builds up
sufficiently over the integration interval to overcome any fluctuations
in $X(s)$, we then have a continuous shrinking of the phase-space of
perturbed dynamics, indicating synchronization.  Integrating 
the exponential term in Eq. (\ref{eq:W}) over many oscillations and using
Eq. (\ref{eq:insert}), we get,
\begin{equation}
 \int^t_{t_1} \{X(s) - \epsilon \left(a_1 + b_1 Y(s) \right) \} \cdot ds  
\approx - \epsilon \left(a_1 + b_1 \bar Y \right) \left(t-t_1\right) + 
\int^t_{t_1+nT} X(s) \cdot ds
\label{eq:insert2}
\end{equation}
where $t-t_1$ is the total integration interval,  $\bar{Y}$ is the average
value of intensity over that interval, and $n$ on the integration
limits is the total integer amount of full oscillations that fit into the
integration period: $t-t_1-nT < T$.  
Here, $T$ is the average period of oscillation
over the integration interval.
Thus the integral of $X(s)$ on the right hand side is
only over a single uncompleted oscillation.  This integral, however,
may still be significant compared to the $\epsilon$ term, 
since its fluctuations are comparable to $X(s)$, 
because the integration period,
$T$, is of order unity (due to scaling in the equations), while $\epsilon$
multiplying the other term is small.  It follows that sufficiently long
integration times, $t-t_1$, 
are required in order for the dissipation term to dominate.
We can now set an upper bound for the integral on the right-hand
side of Equation (\ref{eq:insert2}), 
\begin{equation}
\int^t_{t_1+nT} X(s) \cdot ds < \pi |X(t)|_{max}
\label{eq:bound}
\end{equation}
where $|X(t)|_{max}$ is the maximum fluctuation of 
inversion over the interval of twice the delay time.  The above bound may not
be valid at energies far above the threshold,
when the laser behavior becomes pulse-like with
a period that is significantly longer than the scaled relaxation period of $2
\pi$.  This may be another reason why there is a loss of synchronization
at higher coupling strengths, which lead to higher amplitudes of oscillation,
with lower frequencies.
Requiring Eq. (\ref{eq:insert2}) to be less than zero and using
Eq. (\ref{eq:bound}), we can now obtain a bound above which
the dynamics tend toward the synchronization manifold over
the interval of twice the delay time,
\begin{equation}
 \frac{2}{\pi} \tau \epsilon \left(a_1 + b_1 \bar{Y} \right) > |X(t)|_{max}
\label{eq:compare}
\end{equation}
where we have used $2 \tau = t-t_1$ for the integration interval.  The above
inequality insures the right-hand side of Eq. (\ref{eq:insert2}) is negative
over the interval of twice the delay time.  This in turn 
ensures the shrinking volume of transverse
phase-space dynamics given by Equation (\ref{eq:W}).

Equation (\ref{eq:compare}) gives a condition for
the phase-space volume of transverse dynamics to contract over the interval of
twice the delay.   
For sufficiently long delays, where $\tau \epsilon \left(a_1 + b_1 \bar{Y}
\right) \gg |X(t)|_{max}$, Eq. (\ref{eq:W}) can be approximated as 
\begin{equation}
\ln\left(W(t)\right) = \ln|\triangle x \dot{\triangle y} - \triangle y
\dot{\triangle x}| \approx  - \int^t_{t_1}  \epsilon \left(a_1 + b_1 Y(s) \right) \cdot ds .
\label{eq:contract}
\end{equation}
where a natural log of $W(t)$ was taken.
The above equation
is a monotonically decreasing function of $t$.  This
means that the phase-space volume of the system perturbed from the
synchronization manifold contracts as a function of time.
Applying  Eq. (\ref{eq:LyapSum}) to Eq. (\ref{eq:contract}), where
$\det\left(\mathbf{\Psi(\triangle \mathbf{z_{1,3}})}\right) = W(t)$,
the sum of transverse Lyapunov exponents for the dynamics
off the synchronization manifold described by Eq. (\ref{eq:matrix}) can
now be approximated as, 
\begin{equation}
\lambda_1 + \lambda_2 \approx -\epsilon \left(a_1 + b_1 \bar{Y} \right)
\label{eq:LyapLasers}
\end{equation}
Equation (\ref{eq:LyapLasers}) indicates that for sufficiently long
integration times (requiring sufficiently long delays),
the sum of the Lyapunov exponents should be negative, indicating the shrinking
of phase-space volume of dynamics transverse to the synchronization manifold.

\begin{figure}[ht]
\hspace*{-1.5 cm}
{
\epsfxsize=5in
\epsffile{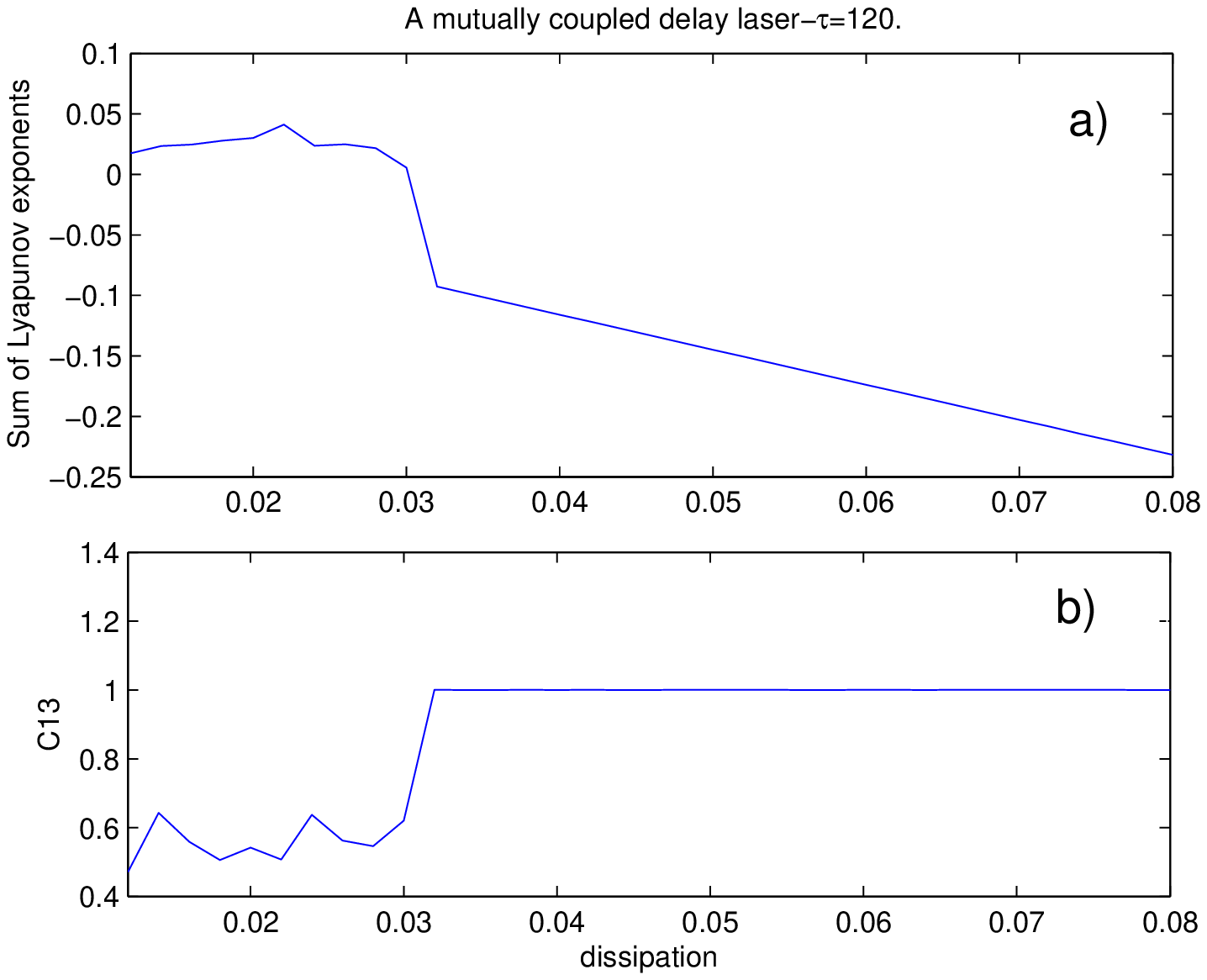}
}
{
\hspace*{-1.3 cm}
\epsfxsize=5in
\epsffile{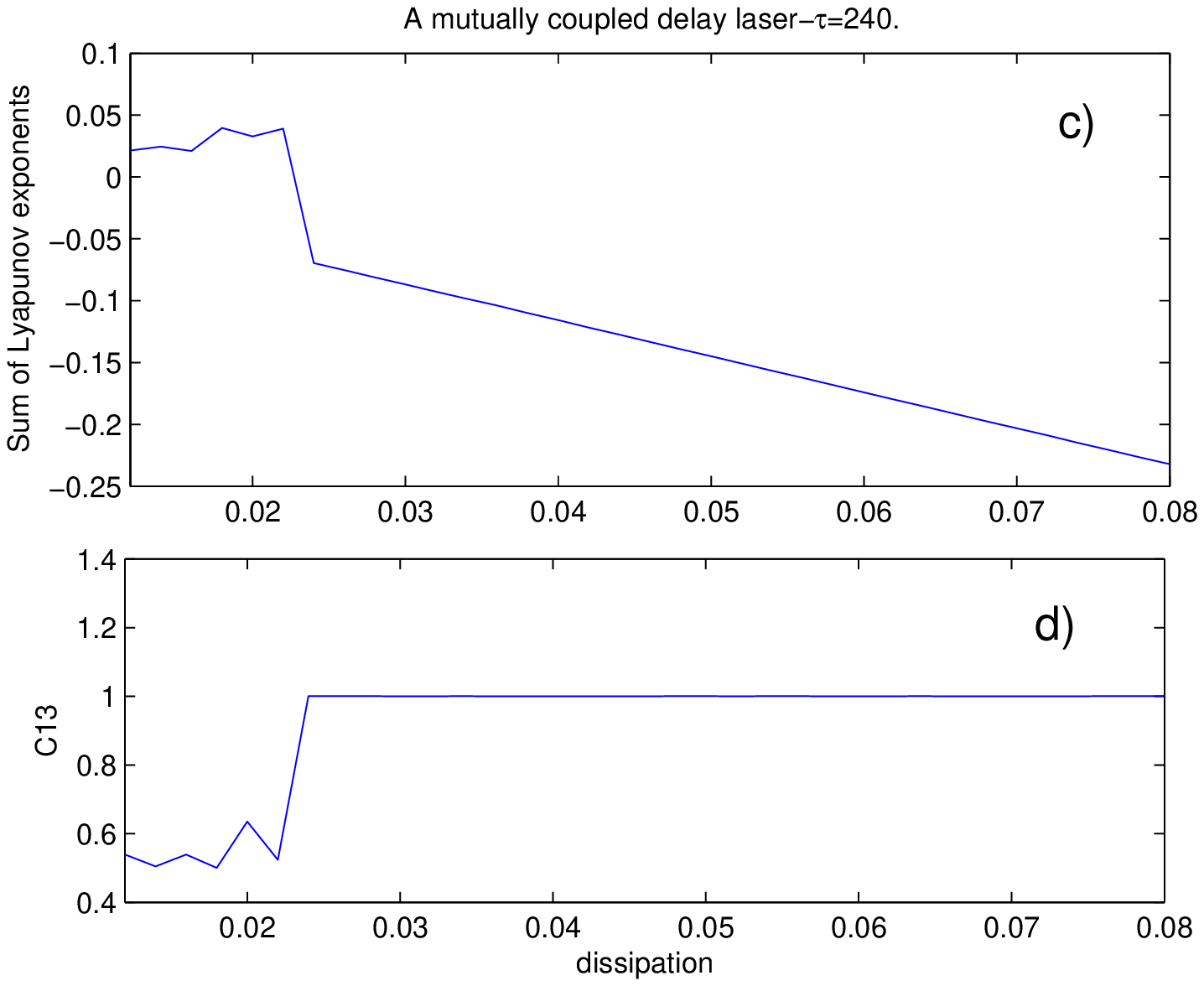}
}
\caption{a) Sum of Lyapunov exponents as a function of dissipation, $\epsilon$,
for $\tau=120$.  b) Corresponding
  correlations between outer lasers, $\tau=120$.  
c) Sum of Lyapunov exponents vs. $\epsilon$,  
for $\tau=240$.  d)  Corresponding
  correlations between outer lasers, $\tau=240$. 
In all cases,
$a_1=a_2=2$, $b_1=b_2=1$, $\delta_1 = \delta_2 = 0.2$, $\beta=0.5$.}
\label{fig:SumLyapEp}
\end{figure}

Figure \ref{fig:SumLyapEp} shows numerically computed sum of
Lyapunov exponents, and corresponding correlations of the outer lasers,
as a function of dissipation, $\epsilon$, for 
two values of the delay, $\tau=120$ and $\tau=240$.  
The fluctuations in the sum of Lyapunov exponents
correspond well to the fluctuations in the correlation function of the outer
lasers, with desynchronization when the Lyapunov sum increases above 
zero.  As might be expected from
Eq. (\ref{eq:compare}), longer delays mean synchronization at lower values of
dissipation, since the dissipation term in the exponential in
Eq. (\ref{eq:W})  dominates for 
sufficiently long delays.  Increasing $\tau$
by a factor of two, however, does not decrease the bifurcation value of
$\epsilon$ for the onset of synchronization by a factor of two, as might be
expected from Eq. (\ref{eq:compare}).  This is probably due to the decrease in
fluctuations, $|X(t)|_{max}$, as the dissipation in the system is increased,
leading to synchronization at a lower value of $\epsilon$ than might otherwise
be expected.  After  Eq. (\ref{eq:compare}) is satisfied, resulting in the
onset of synchronization, the sum of Lyapunov exponents has a negative linear
dependence, given by Eq. (\ref{eq:LyapLasers}).  This is in agreement with
Fig. \ref{fig:SumLyapEp} which shows this negative linear
dependence of Lyapunov sum on dissipation, with a slope of around $-2.6$, a
reasonable value for the parameters used of $a_1=2$, $b_1=1$ and intensity,
$\bar Y \sim 1$.
In general, the average intensity of the dynamics,
$\bar{Y}$, depends on the coupling strengths, $\delta_1$, and $\delta_2$.

While
Eqs. (\ref{eq:contract})  and (\ref{eq:LyapLasers}) predict the shrinking
of phase space for the dynamics transverse to the synchronization manifold,  
to guarantee stability both Lyapunov exponents have to be negative,
or the solution will blow up along the unstable direction.
To find out whether the synchronous state is stable,  consider again the
Wronskian, $W(t) =  \triangle x \dot{\triangle y} 
- \triangle y \dot{\triangle x}$. 
Substituting for $\dot{\triangle x}$ and $\dot{\triangle y}$
from Eq. (\ref{eq:matrix}), we get
\begin{equation}
W(t) = \{1+Y(t)\} \cdot \left(\triangle x\right)^2 + \{1+\epsilon b_1 X(t)\} \cdot \left(\triangle
y\right)^2 + \{ \epsilon \lbrack a_1 + b_1 Y(t) \rbrack + X(t) \} \cdot
\triangle x \triangle y 
\label{eq:Wcontract}
\end{equation}
For $|\epsilon b_1 X(t)| < 1$ (a reasonable assumption since $\epsilon \ll
1$), terms quadratic in $\triangle x$ and $\triangle y$ are always positive,
indicating rotation.
In Eq. (\ref{eq:contract}), $W(t)$ is a monotonically decreasing function of
time, with $W(t) \rightarrow 0$ as $t \rightarrow \infty$.  
Therefore both $\triangle x, \triangle y \rightarrow 0$ as $t
\rightarrow \infty$, due to the presence of positive
terms quadratic in  $\triangle x$ 
and $\triangle y$ in Eq. (\ref{eq:Wcontract}).  It follows that for
sufficiently long delays in the system, the synchronized state is
stable, and therefore all the Lyapunov exponents transverse to
the synchronization manifold are negative.  
The stability of synchronized
state is due to cross-terms in the matrix in Eq. (\ref{eq:matrix}), which come
from rotation, leading to the nonlinear exchange of energy between
inversion and intensity of the laser.  This rotation introduces
positive quadratic terms in $\triangle x$
and $\triangle y$ into Eq.  (\ref{eq:Wcontract}) and leads to the
spiraling of the phase-space volume towards zero, rather that blowing up
along one direction, while shrinking along another.  

\begin{figure}[ht]
\hspace*{-1 cm}
{
\epsfxsize=5in
\epsffile{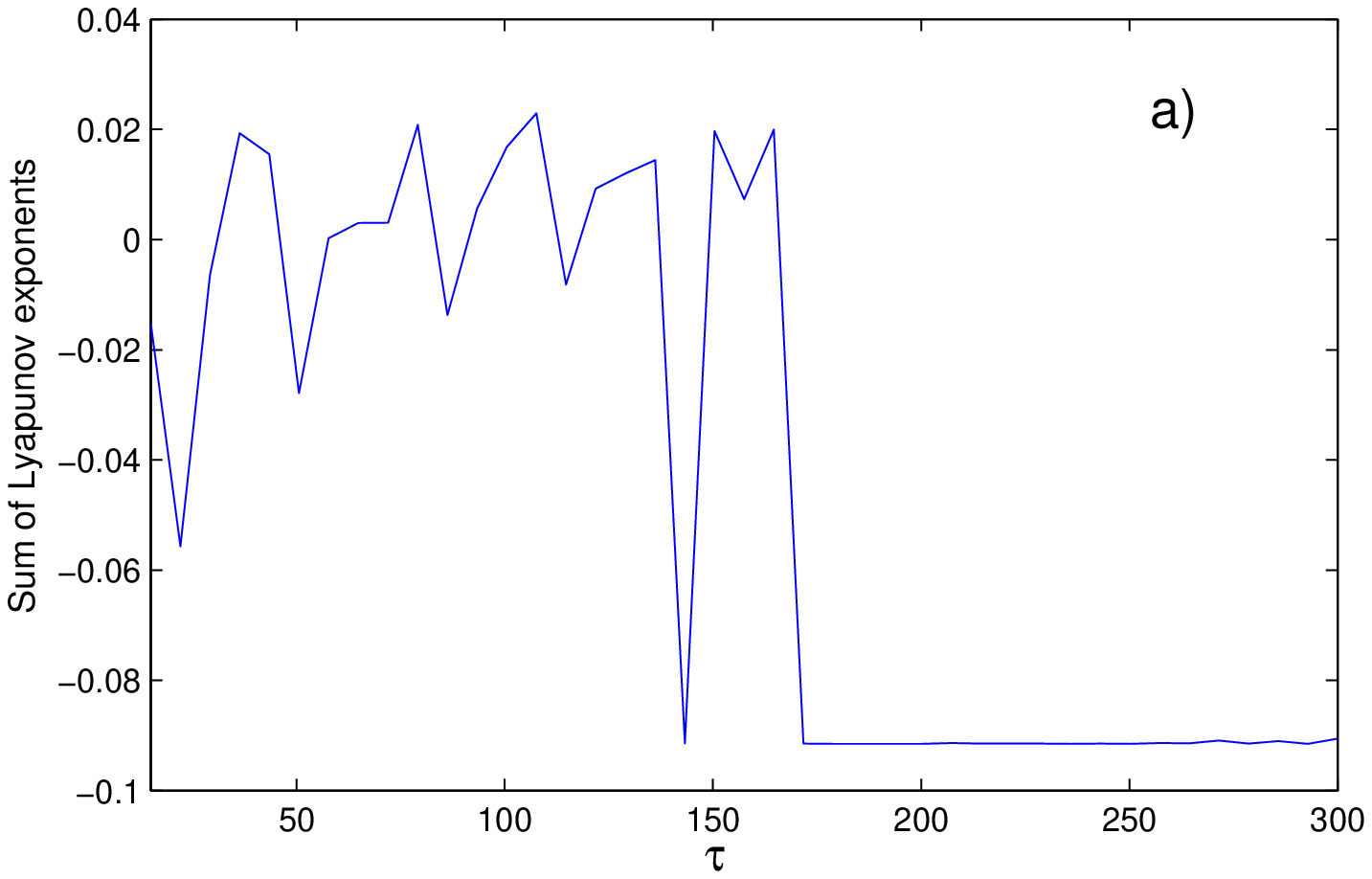}
}
{
\hspace*{-0.25 cm}
\epsfxsize=5in
\epsffile{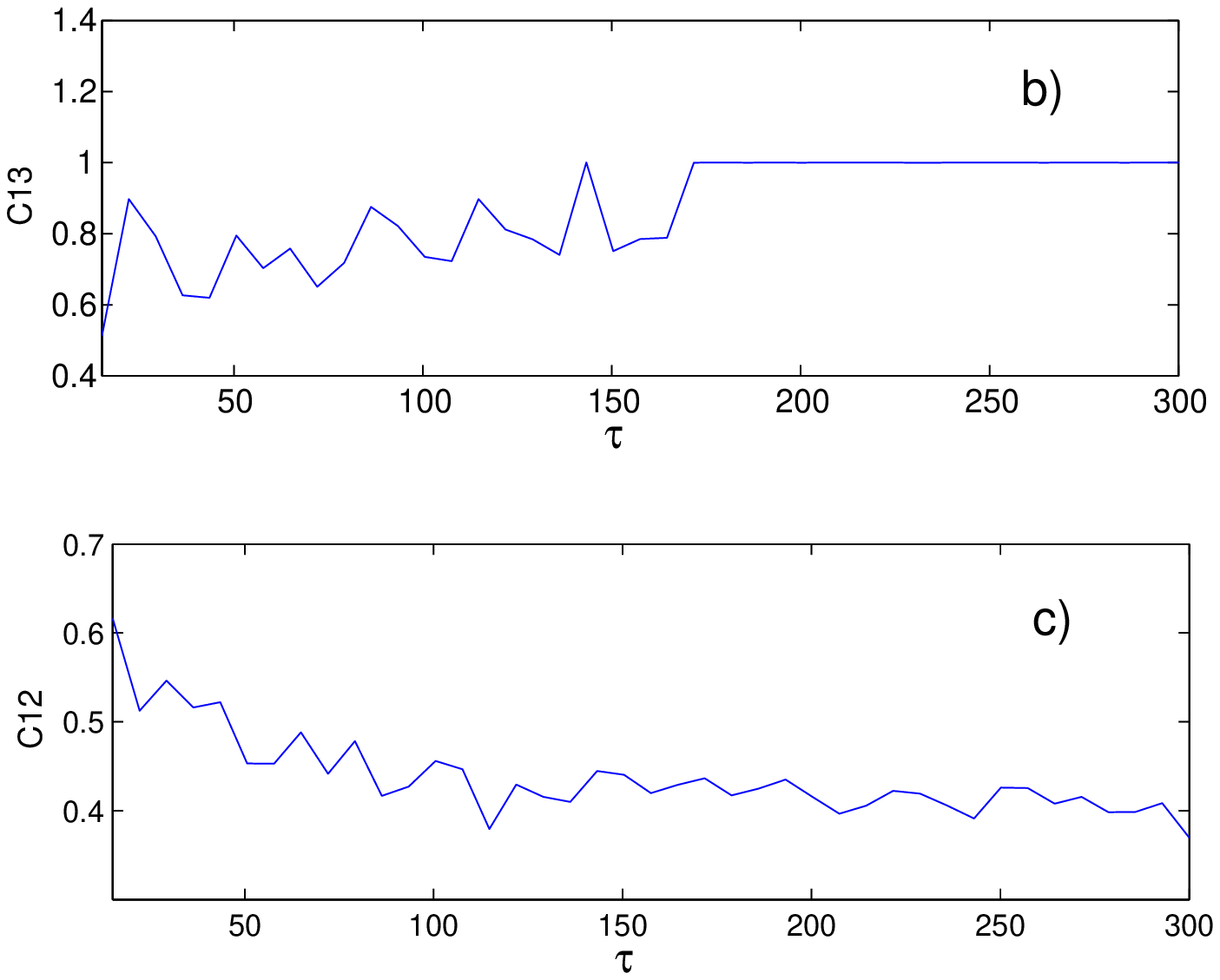}
}
\caption{a) Numerically computed sum of Lyapunov exponents as 
a function of delay, $\tau$.  b) Corresponding correlations of outer lasers.
c) Correlations of the middle and outer lasers, shifted by
the delay time to maximize correlations.  
$\epsilon = \sqrt{0.001}$, $\delta_1=\delta_2=7.5 \epsilon$, $\beta=0.5$.}
\label{fig:SumLyapDelay}
\end{figure}

Figure  \ref{fig:SumLyapDelay}a
shows the sum of Lyapunov exponents as a function of delay.  
The Lyapunov exponents are negative for all $\tau > 170$, (corresponding 
to about $60$ ns) resulting in
complete synchronization of the outer lasers, as shown in Fig.
\ref{fig:SumLyapDelay}b.  
At the same time, the outer lasers are not synchronized with the
center one, Fig. \ref{fig:SumLyapDelay}c.  The fluctuations in correlations
of the outer lasers match well the fluctuations in the Lyapunov sum, with
correlations increasing whenever the Lyapunov sum decreases.  
Figure  \ref{fig:SumLyapDelay} agrees well with the above analysis, since
sufficiently long delays (see Eq. (\ref{eq:compare}))
are needed for the Lyapunov exponents to become
negative, leading to synchronization.  After the onset of synchronization,
Eq. (\ref{eq:LyapLasers}) becomes valid, so that the Lyapunov sum
becomes independent of delays.  This is confirmed by the straight horizontal
line in the figure, after the outer lasers synchronize. The degree of
synchronization is given by the correlations function. 

From Eq. (\ref{eq:LyapLasers}), the negative Lyapunov exponents, leading to
stability of synchronous state, are the result of
dissipation, $\epsilon$, in the end lasers.  
This is to be expected since mutual
coupling pumps energy into the system, as can be seen in
Eqs. (\ref{eq:lasers1}) - (\ref{eq:lasers}).  Therefore, some dissipation in
the outer subsystems themselves is essential in order to ``wash out'' their
dependence on initial conditions and make them a function of the dynamics of
the middle laser, as would be required in the case on complete
synchronization.  

\begin{figure}[ht]
\hspace*{-1 cm}
{
\epsfxsize=6in
\epsffile{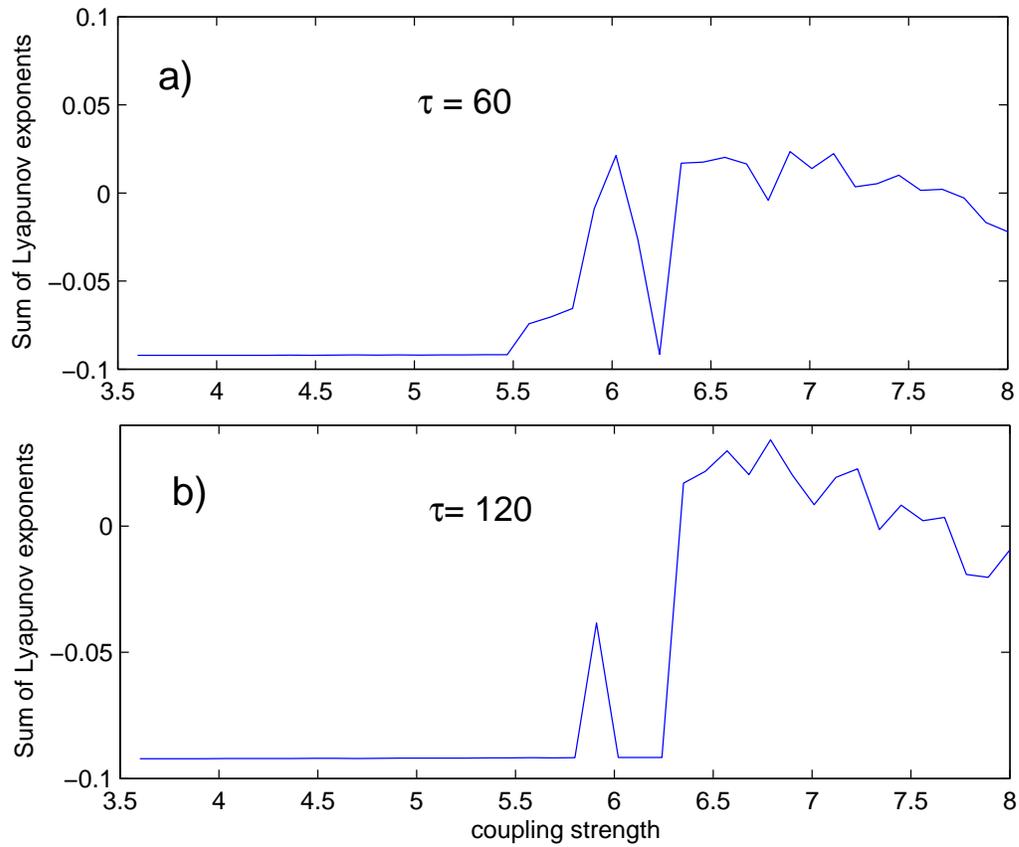}
}
\caption{a) Sum of Lyapunov exponents as 
a function of coupling strength, $\delta_1=\delta_2$, for $\tau=60$.  
b) $\tau=120$. 
$\epsilon = \sqrt{0.001}$, $\beta=0.5$.}
\label{fig:SumLyapCoupling}
\end{figure}

\begin{figure}[ht]
\hspace*{-1 cm}
{
\epsfxsize=5in
\epsffile{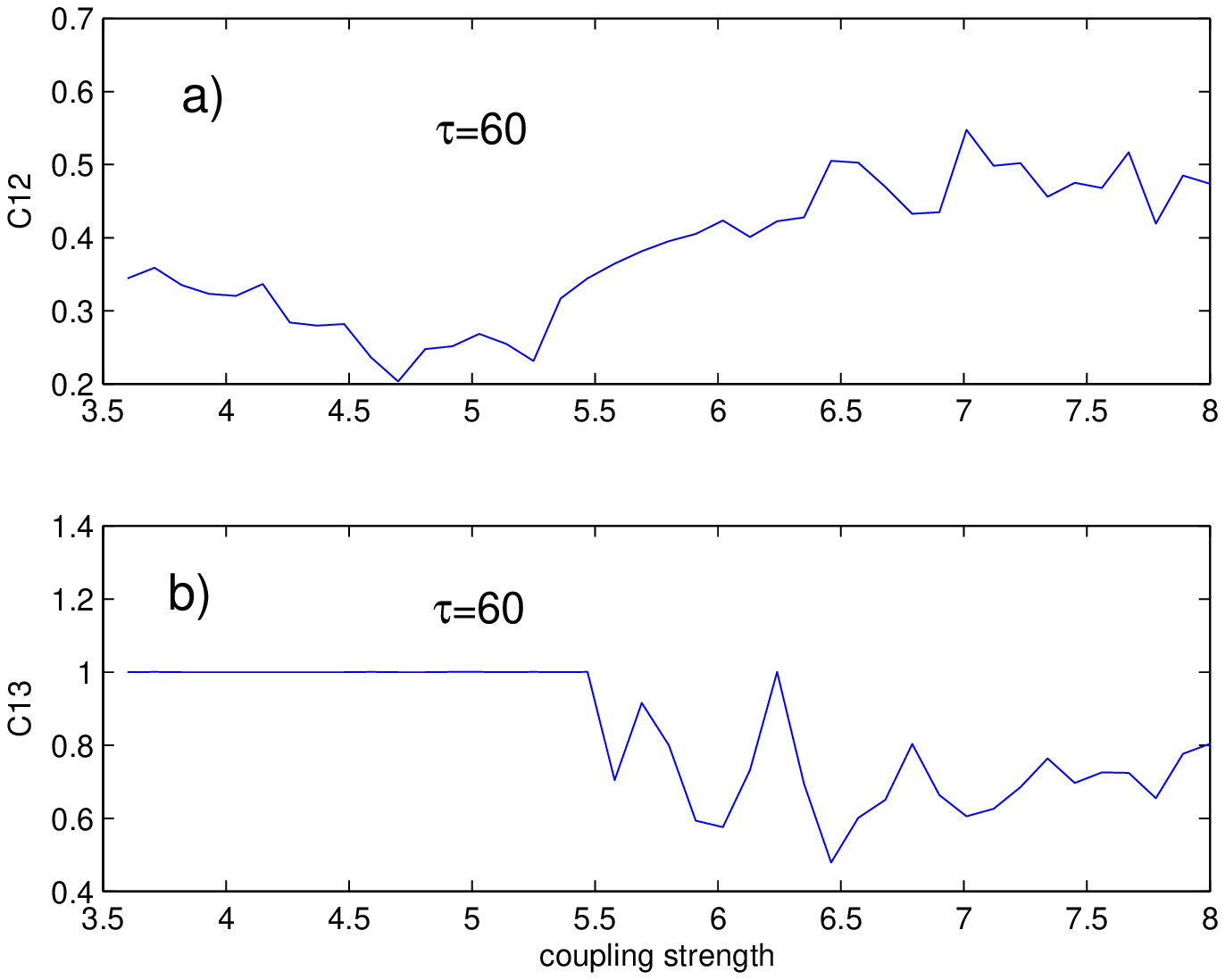}
}
{
\hspace*{-0.7 cm}
\epsfxsize=5in
\epsffile{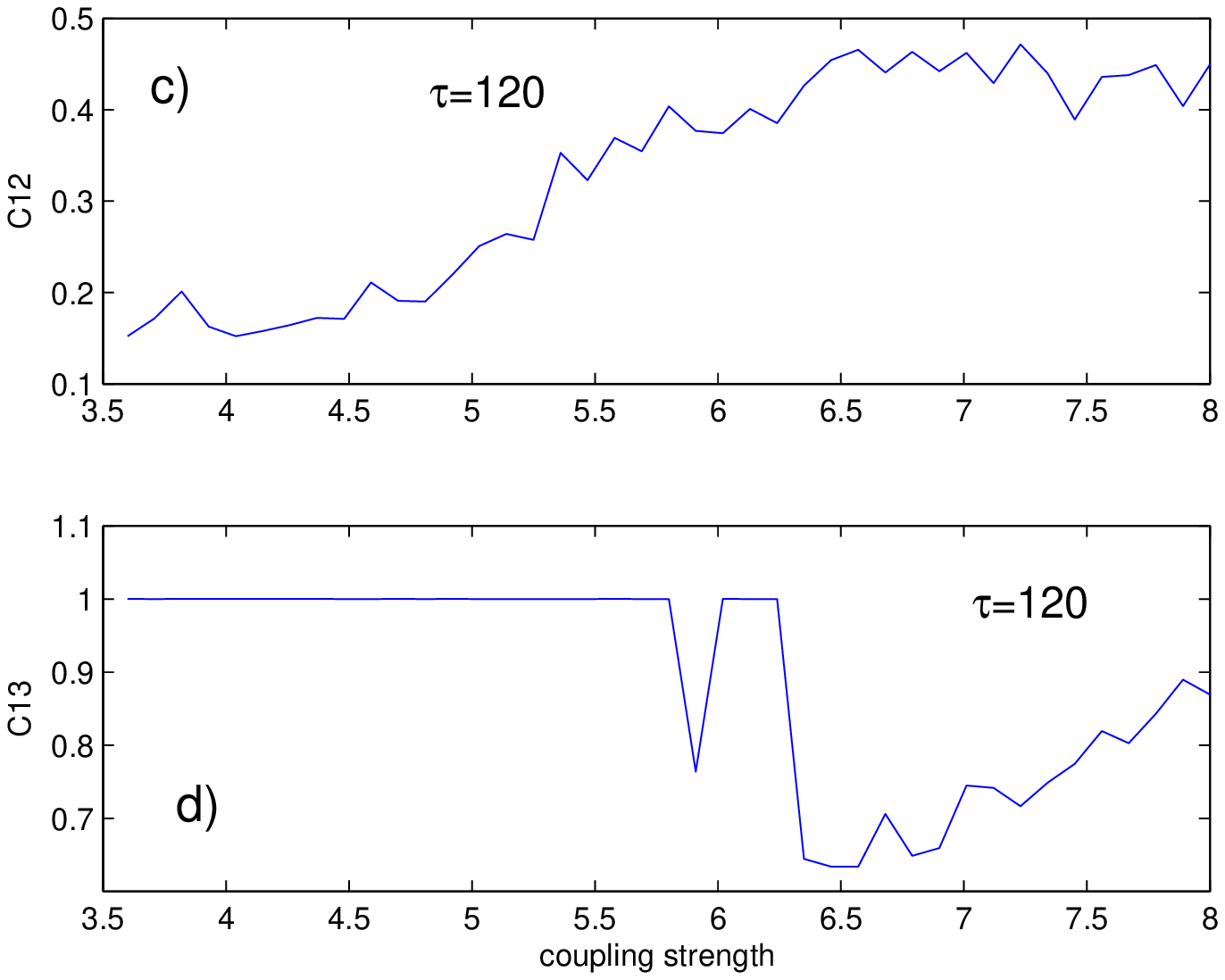}
}
\caption{Correlations corresponding to Fig. \ref{fig:SumLyapCoupling}.
a)  Correlation between the middle and one of the outer lasers, $\tau=60$.
b) Correlations of outer lasers, $\tau=60$.
c)  Correlation between the middle and one of the outer lasers, $\tau=120$.
d) Correlations of outer lasers, $\tau=120$.  
Outer lasers synchronize for greater range of
coupling strength as the delay is increased.  The middle and the outer
lasers show little correlation for all values of the coupling strengths.}
\label{fig:CorrCoupling}
\end{figure}

The amplitude of laser oscillations 
depends on the coupling strengths, $\delta_1$
and $\delta_2$, as well as the dissipation. It was shown \cite{KimRACS05}
that in the case of a two laser system there is a bifurcation value for the
onset of oscillations that is a function of product of the coupling
strengths, $\delta_1 \delta_2$.  Increasing the coupling strengths increases
the role of nonlinearities in the system and the intensity of 
laser oscillations.
Thus for low values of the coupling strengths, the dynamics given by
Eqs. (\ref{eq:lasers1}) - (\ref{eq:lasers}) are more regular.  At low
intensities, the dynamics of individual lasers are close to that of a simple
harmonic oscillator, as can be verified by substituting low values of
$\{x_1,y_1\}$ into Eq. (\ref{eq:lasers1}), for example.  In order for
Eq. (\ref{eq:corr}) to be valid, the dynamics have to be uncorrelated over the
time interval of the delay.  Thus 
the equations derived in this section are
valid for chaotic regime which requires sufficiently high product of
coupling strengths, $\delta_1 \delta_2$.  In this case, it can be assumed that
the outer lasers are driven by an uncorrelated signal from the middle
laser over the time interval of $2 \tau$.
Any synchronization on that interval would then be analogous to generalized
synchronization that occurs in a uni-directional system, with the middle
laser acting as the driver for the outer ones.  
Since increased coupling pumps
more energy into the system, thereby increasing the effect of 
nonlinearities, the Lyapunov exponents may increase above zero, 
leading to desynchronization of the outer lasers.  In this
case, longer delays in coupling may be required in order for the outer lasers
to synchronize.  This effect is illustrated in Fig. \ref{fig:SumLyapCoupling},
which shows the sum of Lyapunov exponents as a function of coupling strengths
for two different delays, $\tau=60$ and $\tau=120$.  There is an 
abrupt increase in Lyapunov exponents above zero, due to increased
nonlinearity, as the coupling strength is increased. 
Increasing the delay however to $\tau=120$ leads to
synchronization for a greater range of coupling strengths, as compared
to $\tau=60$.  The corresponding correlations as a function of coupling
strengths are shown in Fig. \ref{fig:CorrCoupling}.  
Desynchronization at higher coupling strengths, and the synchronizing effect
of increased delays is in agreement with Eq. (\ref{eq:compare}).  
Since higher coupling strengths lead to greater fluctuations in
 $X(t)$, longer integration times are required in order for
Eqs. (\ref{eq:contract}) and (\ref{eq:LyapLasers})  to be valid, 
leading to synchronization at longer delays, $\tau$.  

Figures \ref{fig:SumLyapEp} - \ref{fig:CorrCoupling} show that Eq.
(\ref{eq:LyapLasers}) correctly predicts the independence of Lyapunov
sum on delays and coupling strengths and a negative linear dependence on
dissipation, once synchronization sets in.  Synchronization, on the other hand,
occurs once the condition expressed in Eq. (\ref{eq:compare}) is satisfied,
leading to the continuous shrinking of the phase-space dynamics transverse to
the synchronization manifold.


\section{\label{sec:concl} IV. Conclusion}
Ideas from generalized
synchronization were used to explain complete chaotic
synchronization of mutually coupled systems in the presence of long delays.
Since identical outer subsystems synchronize due to a common input
from the middle subsystem, complete synchronization is similar
to the one occurring in the auxiliary system set-up, with the exception
that all subsystems are mutually coupled.  This leads to the 
dependence of common input to the outer subsystems 
on history of the dynamics.  
Complete chaotic synchronization is the result
of the outer systems becoming a function of the middle one, as would
happen in the case of generalized synchronization.

Due to the symmetry of the outer subsystems, the dynamics  linearized about the synchronization
manifold are independent of explicit coupling.  Transverse Lyapunov 
exponents can then be calculated to determine
the stability of the synchronous state.  Since over the time scale of
twice the delay interval, the outer subsystems can be viewed as driven
by a common
chaotic signal from the middle subsystem, the analysis is considerably
simplified, allowing for calculation of phase-space volume dynamics
transverse to the synchronization manifold.   
The transverse phase-space volume dynamics were analyzed for the case of
three mutually coupled semiconductor lasers.  It was found that for
sufficiently long delays, the synchronized state is stable.  The
sum of Lyapunov exponents transverse to the synchronization manifold
was found analytically and shown to have a negative linear dependence
on dissipation, in good agreement with numerical calculations.  
This also confirmed the
intuition that synchronization is the result of dissipation,
$\epsilon$ in local dynamics of the
lasers themselves, since the coupling between lasers is not dissipative.
The analysis also explains the effect of various parameters on
synchronization, such as coupling strengths, delay time, and 
dissipation, and is supported by numerical simulations over a range of
parameter values.  Thus, it was shown analytically and confirmed
numerically that after the onset of synchronization, the stability of the
synchronous state (as given by Lyapunov exponents) depends linearly on
dissipation, but is independent of the delay time and coupling strength.

\section{\label{sec:ack} V. Acknowledgments}
Valuable discussions with Louis Pecora are gratefully acknowledged.
The authors thank the Office of naval Research for their continued
support in the research presented. ASL is currently a National
Research Council post doctoral fellow.


\begin{thebibliography}{18}
\expandafter\ifx\csname natexlab\endcsname\relax\def\natexlab#1{#1}\fi
\expandafter\ifx\csname bibnamefont\endcsname\relax
  \def\bibnamefont#1{#1}\fi
\expandafter\ifx\csname bibfnamefont\endcsname\relax
  \def\bibfnamefont#1{#1}\fi
\expandafter\ifx\csname citenamefont\endcsname\relax
  \def\citenamefont#1{#1}\fi
\expandafter\ifx\csname url\endcsname\relax
  \def\url#1{\texttt{#1}}\fi
\expandafter\ifx\csname urlprefix\endcsname\relax\def\urlprefix{URL }\fi
\providecommand{\bibinfo}[2]{#2}
\providecommand{\eprint}[2][]{\url{#2}}

\bibitem[{\citenamefont{Fischer et~al.}(2006)\citenamefont{Fischer, Vicente,
  Buldu, Peil, Mirasso, Torrent, and Garcia-Ojalvo}}]{FishcervbpMtG06}
\bibinfo{author}{\bibfnamefont{I.}~\bibnamefont{Fischer}},
  \bibinfo{author}{\bibfnamefont{R.}~\bibnamefont{Vicente}},
  \bibinfo{author}{\bibfnamefont{J.~M.} \bibnamefont{Buldu}},
  \bibinfo{author}{\bibfnamefont{M.}~\bibnamefont{Peil}},
  \bibinfo{author}{\bibfnamefont{C.~R.} \bibnamefont{Mirasso}},
  \bibinfo{author}{\bibfnamefont{M.~C.} \bibnamefont{Torrent}}, \bibnamefont{and}
  \bibinfo{author}{\bibfnamefont{J.}~\bibnamefont{Garcia-Ojalvo}},
  \bibinfo{journal}{Physical Review Letters} \textbf{\bibinfo{volume}{97}},
  \bibinfo{pages}{123902} (\bibinfo{year}{2006}).

\bibitem[{\citenamefont{Mirasso et~al.}(2004)\citenamefont{Mirasso, Vicente,
  Colet, Mulet, and Perez}}]{MirassoVCMP04}
\bibinfo{author}{\bibfnamefont{C.~R.} \bibnamefont{Mirasso}},
  \bibinfo{author}{\bibfnamefont{R.}~\bibnamefont{Vicente}},
  \bibinfo{author}{\bibfnamefont{P.}~\bibnamefont{Colet}},
  \bibinfo{author}{\bibfnamefont{J.}~\bibnamefont{Mulet}}, \bibnamefont{and}
  \bibinfo{author}{\bibfnamefont{T.}~\bibnamefont{Perez}},
  \bibinfo{journal}{Comptes Rendus Physique} \textbf{\bibinfo{volume}{5}},
  \bibinfo{pages}{613} (\bibinfo{year}{2004}).

\bibitem[{\citenamefont{Ciszak et~al.}(2003)\citenamefont{Ciszak, Calvo,
  Masoller, Mirasso, and Toral}}]{CiszakCMMT03}
\bibinfo{author}{\bibfnamefont{M.}~\bibnamefont{Ciszak}},
  \bibinfo{author}{\bibfnamefont{O.}~\bibnamefont{Calvo}},
  \bibinfo{author}{\bibfnamefont{C.}~\bibnamefont{Masoller}},
  \bibinfo{author}{\bibfnamefont{C.~R.} \bibnamefont{Mirasso}},
  \bibnamefont{and} \bibinfo{author}{\bibfnamefont{R.}~\bibnamefont{Toral}},
  \bibinfo{journal}{Physical Review Letters} \textbf{\bibinfo{volume}{90}},
  \bibinfo{pages}{204102} (\bibinfo{year}{2003}).

\bibitem[{\citenamefont{Pikovsky et al.}(2001)\citenamefont{Pikovsky,
  Rosenblum, and Kurths}}]{pikovsky}
\bibinfo{author}{\bibfnamefont{A.}~\bibnamefont{Pikovsky}},
  \bibinfo{author}{\bibfnamefont{M.}~\bibnamefont{Rosenblum}},
  \bibnamefont{and} \bibinfo{author}{\bibfnamefont{J.}~\bibnamefont{Kurths}},
  \emph{\bibinfo{title}{Synchronization: A universal concept in nonlinear
  science}} (\bibinfo{publisher}{Cambridge University Press},
  \bibinfo{address}{Cambridge}, \bibinfo{year}{2001}).

\bibitem[{\citenamefont{Boccaletti et~al.}(2002)\citenamefont{Boccaletti,
  Kurths, Osipov, Valladares, and Zhou}}]{BoccalettiKOVZ02}
\bibinfo{author}{\bibfnamefont{S.}~\bibnamefont{Boccaletti}},
  \bibinfo{author}{\bibfnamefont{J.}~\bibnamefont{Kurths}},
  \bibinfo{author}{\bibfnamefont{G.}~\bibnamefont{Osipov}},
  \bibinfo{author}{\bibfnamefont{D.~L.} \bibnamefont{Valladares}},
  \bibnamefont{and} \bibinfo{author}{\bibfnamefont{C.~S.} \bibnamefont{Zhou}},
  \bibinfo{journal}{Physics Reports-Review Section Of Physics Letters}
  \textbf{\bibinfo{volume}{366}}, \bibinfo{pages}{1} (\bibinfo{year}{2002}).

\bibitem[{\citenamefont{Hale}(1971)}]{Hale:1971:FDE}
\bibinfo{author}{\bibfnamefont{J.}~\bibnamefont{Hale}},
  \emph{\bibinfo{title}{Functional Differential Equations}}
  (\bibinfo{publisher}{Springer- Verlag}, \bibinfo{address}{New York},
  \bibinfo{year}{1971}).

\bibitem[{\citenamefont{Pecora and Carroll}(1998)}]{PecoraC98}
\bibinfo{author}{\bibfnamefont{L.~M.} \bibnamefont{Pecora}} \bibnamefont{and}
  \bibinfo{author}{\bibfnamefont{T.~L.} \bibnamefont{Carroll}},
  \bibinfo{journal}{Physical Review Letters} \textbf{\bibinfo{volume}{80}},
  \bibinfo{pages}{2109} (\bibinfo{year}{1998}).

\bibitem[{\citenamefont{Terry et~al.}(1999)\citenamefont{Terry, Thornburg,
  DeShazer, Vanwiggeren, Zhu, Ashwin, and Roy}}]{TerryTDVZAR99}
\bibinfo{author}{\bibfnamefont{J.~R.} \bibnamefont{Terry}},
  \bibinfo{author}{\bibfnamefont{K.~S.} \bibnamefont{Thornburg}},
  \bibinfo{author}{\bibfnamefont{D.~J.} \bibnamefont{DeShazer}},
  \bibinfo{author}{\bibfnamefont{G.~D.} \bibnamefont{Vanwiggeren}},
  \bibinfo{author}{\bibfnamefont{S.} \bibnamefont{Zhu}},
  \bibinfo{author}{\bibfnamefont{P.}~\bibnamefont{Ashwin}}, \bibnamefont{and}
  \bibinfo{author}{\bibfnamefont{R.}~\bibnamefont{Roy}},
  \bibinfo{journal}{Physical Review E} \textbf{\bibinfo{volume}{59}},
  \bibinfo{pages}{4036} (\bibinfo{year}{1999}).

\bibitem[{\citenamefont{Carr et~al.}(2006{\natexlab{a}})\citenamefont{Carr,
  Taylor, and Schwartz}}]{CarrTS06}
\bibinfo{author}{\bibfnamefont{T.~W.} \bibnamefont{Carr}},
  \bibinfo{author}{\bibfnamefont{M.~L.} \bibnamefont{Taylor}},
  \bibnamefont{and} \bibinfo{author}{\bibfnamefont{I.~B.}
  \bibnamefont{Schwartz}}, \bibinfo{journal}{Physica D-Nonlinear Phenomena}
  \textbf{\bibinfo{volume}{213}}, \bibinfo{pages}{152}
  (\bibinfo{year}{2006}{\natexlab{a}}).

\bibitem[{not()}]{note3}
\bibinfo{note}{No assumptions are made about the local dynamics possessing any
  chaotic attractors. For the $N=3$ laser case, each uncoupled laser only has a
  unique steady state. That is, only non-trivial behavior can be induced either
  by external forcing, self feedback with delays, or by coupling to other
  lasers.}

\bibitem[{\citenamefont{Rulkov et~al.}(1995)\citenamefont{Rulkov, Sushchik,
  Tsimring, and Abarbanel}}]{RulkovSTA95}
\bibinfo{author}{\bibfnamefont{N.~F.} \bibnamefont{Rulkov}},
  \bibinfo{author}{\bibfnamefont{M.~M.} \bibnamefont{Sushchik}},
  \bibinfo{author}{\bibfnamefont{L.~S.} \bibnamefont{Tsimring}},
  \bibnamefont{and} \bibinfo{author}{\bibfnamefont{H.~D.~I.}
  \bibnamefont{Abarbanel}}, \bibinfo{journal}{Physical Review E}
  \textbf{\bibinfo{volume}{51}}, \bibinfo{pages}{980} (\bibinfo{year}{1995}).

\bibitem[{\citenamefont{Abarbanel et~al.}(1996)\citenamefont{Abarbanel, Rulkov,
  and Sushchik}}]{AbarbanelRS96}
\bibinfo{author}{\bibfnamefont{H.~D.~I.} \bibnamefont{Abarbanel}},
  \bibinfo{author}{\bibfnamefont{N.~F.} \bibnamefont{Rulkov}},
  \bibnamefont{and} \bibinfo{author}{\bibfnamefont{M.~M.}
  \bibnamefont{Sushchik}}, \bibinfo{journal}{Physical Review E}
  \textbf{\bibinfo{volume}{53}}, \bibinfo{pages}{4528} (\bibinfo{year}{1996}).

\bibitem[{\citenamefont{Hartman}(1982)}]{Hartman:1982:ODE}
\bibinfo{author}{\bibfnamefont{P.}~\bibnamefont{Hartman}},
  \emph{\bibinfo{title}{Ordinary Differential Equations}}
  (\bibinfo{publisher}{Birkh{\"{a}}user}, \bibinfo{address}{Boston},
  \bibinfo{year}{1982}), \bibinfo{edition}{2nd} ed., ISBN
  \bibinfo{isbn}{3-7643-3068-6}.

\bibitem[{\citenamefont{Guan et~al.}(2006)\citenamefont{Guan, Lai, Lai, and
  Gong}}]{GuanLLG06}
\bibinfo{author}{\bibfnamefont{S.}~\bibnamefont{Guan}},
  \bibinfo{author}{\bibfnamefont{Y.}~\bibnamefont{Lai}},
  \bibinfo{author}{\bibfnamefont{C.}~\bibnamefont{Lai}}, \bibnamefont{and}
  \bibinfo{author}{\bibfnamefont{X.}~\bibnamefont{Gong}},
  \bibinfo{journal}{Physics Letters A} \textbf{\bibinfo{volume}{353}},
  \bibinfo{pages}{30} (\bibinfo{year}{2006}).

\bibitem[{\citenamefont{Kim et~al.}(2005)\citenamefont{Kim, Roy, Aron, Carr,
  and Schwartz}}]{KimRACS05}
\bibinfo{author}{\bibfnamefont{M.~Y.} \bibnamefont{Kim}},
  \bibinfo{author}{\bibfnamefont{R.}~\bibnamefont{Roy}},
  \bibinfo{author}{\bibfnamefont{J.~L.} \bibnamefont{Aron}},
  \bibinfo{author}{\bibfnamefont{T.~W.} \bibnamefont{Carr}}, \bibnamefont{and}
  \bibinfo{author}{\bibfnamefont{I.~B.} \bibnamefont{Schwartz}},
  \bibinfo{journal}{Physical Review Letters} \textbf{\bibinfo{volume}{94}},
  \bibinfo{pages}{Art. no. 088101} (\bibinfo{year}{2005}).

\bibitem[{\citenamefont{Carr et~al.}(2006{\natexlab{b}})\citenamefont{Carr,
  Schwartz, Kim, and Roy}}]{carrS06}
\bibinfo{author}{\bibfnamefont{T.~W.} \bibnamefont{Carr}},
  \bibinfo{author}{\bibfnamefont{I.~B.} \bibnamefont{Schwartz}},
  \bibinfo{author}{\bibfnamefont{M.-Y.} \bibnamefont{Kim}}, \bibnamefont{and}
  \bibinfo{author}{\bibfnamefont{R.}~\bibnamefont{Roy}}
  (\bibinfo{year}{2006}{\natexlab{b}}),
  \bibinfo{note}{http://arxiv.org/abs/nlin.CD/0608015}.

\bibitem[{\citenamefont{Schwartz and Erneux}(1994)}]{SchwartzE94}
\bibinfo{author}{\bibfnamefont{I.~B.} \bibnamefont{Schwartz}} \bibnamefont{and}
  \bibinfo{author}{\bibfnamefont{T.}~\bibnamefont{Erneux}},
  \bibinfo{journal}{Siam Journal On Applied Mathematics}
  \textbf{\bibinfo{volume}{54}}, \bibinfo{pages}{1083} (\bibinfo{year}{1994}).

\bibitem[{\citenamefont{Vicente et~al.}(2004)\citenamefont{Vicente, Tang,
      Mulet, Mirasso, and Liu}}]{Vicente04}
\bibinfo{author}{\bibfnamefont{R.}~\bibnamefont{Vicente}},
  \bibinfo{author}{\bibfnamefont{S.}~\bibnamefont{Tang}},
  \bibinfo{author}{\bibfnamefont{J.}~\bibnamefont{Mulet}},
  \bibinfo{author}{\bibfnamefont{C.~R.} \bibnamefont{Mirasso}}, 
\bibnamefont{and}
  \bibinfo{author}{\bibfnamefont{J.~M.} \bibnamefont{Liu}},
  \bibinfo{journal}{Physical Review E} \textbf{\bibinfo{volume}{70}},
  \bibinfo{pages}{046216} (\bibinfo{year}{2004}).

\bibitem[{\citenamefont{Vicente et~al.}(2006)\citenamefont{Vicente, Tang,
      Mulet, Mirasso, and Liu}}]{Vicente06}
\bibinfo{author}{\bibfnamefont{R.}~\bibnamefont{Vicente}},
  \bibinfo{author}{\bibfnamefont{S.}~\bibnamefont{Tang}},
  \bibinfo{author}{\bibfnamefont{J.}~\bibnamefont{Mulet}},
  \bibinfo{author}{\bibfnamefont{C.~R.} \bibnamefont{Mirasso}}, 
\bibnamefont{and}
  \bibinfo{author}{\bibfnamefont{J.~M.} \bibnamefont{Liu}},
  \bibinfo{journal}{Physical Review E} \textbf{\bibinfo{volume}{73}},
  \bibinfo{pages}{047201} (\bibinfo{year}{2006}).

\bibitem[{\citenamefont{Zill}(1982)}]{zill}
\bibinfo{author}{\bibfnamefont{D.}~\bibnamefont{Zill}}, \emph{\bibinfo{title}{A
  first course in differential equations}} (\bibinfo{publisher}{PWS
  publishers}, \bibinfo{address}{Boston}, \bibinfo{year}{1982}).

\end{thebibliography}



\newpage

\end{document}